# Generation and detection of coherent longitudinal acoustic waves in ultrathin 1$T'$-MoTe$_2$

Nicolas Rivas[1], Shazhou Zhong[2], Tina Dekker[2], Meixin Cheng[1], Patrick Gicala[1], Fangchu Chen[3,4], Xuan Luo[3], Yuping Sun[3,5,6], Ariel A. Petruk[1], Kostyantyn Pichugin[1], Adam W. Tsen[2] and Germán Sciaini[1,*]

[1] The Ultrafast electron Imaging Lab, Department of Chemistry, and Waterloo Institute for Nanotechnology, University of Waterloo, Waterloo, ON N2L 3G1, Canada.

[2] Institute for Quantum Computing, Department of Physics and Astronomy, Department of Electrical and Computer Engineering, and Department of Chemistry, University of Waterloo, Waterloo, ON N2L 3G1, Canada.

[3] Key Laboratory of Materials Physics, Institute of Solid-State Physics, Chinese Academy of Sciences, Hefei 230031, People's Republic of China.

[4] University of Science and Technology of China, Hefei, 230026, China.

[5] High Magnetic Field Laboratory, Chinese Academy of Sciences, Hefei, 230031, China

[6] Collaborative Innovation Center of Advanced Microstructures, Nanjing University, Nanjing, 210093, China.

([*]) Correspondence should be addressed to: G. Sciaini, gsciaini@uwaterloo.ca

**ABSTRACT**

Layered transition metal dichalcogenides have attracted substantial attention owing to their versatile functionalities and compatibility with current nanofabrication technologies. Thus, noninvasive means to determine the mechanical properties of nanometer (nm) thick specimens are of increasing importance. Here, we report on the detection of coherent longitudinal acoustic phonon modes generated by impulsive femtosecond (fs) optical excitation. Broadband fs-transient absorption experiments in $1T'$-MoTe$_2$ flakes as a function of thickness (7 nm – 30 nm) yield a longitudinal sound speed of $v_L = (2990 \pm 90)$ m s$^{-1}$. In addition, temperature dependent measurements unveil a linear decrease of the normalized Young's modulus $E_L/E_{L,295\,K}$ with a slope of $\delta(E_L/E_{L,295\,K})/\delta T = (-2.0 \pm 0.1)\,10^{-3}$ K$^{-1}$ and no noticeable change caused by the $T_d - 1T'$ structural phase transition or variations in film thickness.

**MAIN BODY**

The investigation of elastic waves produced by transient surface heating induced by electromagnetic radiation and electron beam bombardment dates back to about 1960[1–3]. The impulsive deposition of energy is known to create a temperature gradient normal to the surface, which leads to thermal stress and the release of elastic waves that propagate away from the heated interface. The detection of acoustic waves in bulk materials was achieved by the implementation of piezoelectric crystals capable of transducing pressure shock waves into readable voltage signals[1–3]. The advent of femtosecond (fs) laser-based techniques made possible the determination of elastic properties in nanomaterials by compensating the drastic reduction in length scale through the improvement in time resolution. When an optical pulse hits the surface of a metallic film for instance, the absorbed photons drive the electron gas out of equilibrium. Fast electron-electron scattering leads to a hotter Fermi distribution, which ultimately transfers the excess energy to the phonon bath on the picosecond timescale. These processes are relatively quick when compared to the thermally-induced macroscopic volume expansion, and therefore electronic and thermal contributions to the stress result in the generation of coherent acoustic waves that locally modulate the structure and the dielectric properties of the material, which can be detected by structural[4–13] and optical probes[14–21].

Here, we implemented fs-broadband transient absorption spectroscopy to generate and detect longitudinal acoustic waves in mechanically exfoliated 1$T'$-MoTe$_2$ crystalline flakes of varying thicknesses (7 nm – 30 nm). Our measurements serve as a means to determine the longitudinal elastic Young's modulus of tiny specimens in a noninvasive manner. Since thin 1$T'$-MoTe$_2$ is known to degrade in the presence of ambient moisture[22], flakes were first transferred onto 10 μm x 10 μm square, 50-nm thick stoichiometric silicon nitride (Si$_3$N$_4$) windows and covered with hexagonal boron nitride (BN) inside a nitrogen-filled glovebox. Furthermore, all experiments were carried out under vacuum conditions in an optical cryostat with the capability to control the temperature from 77 K to 500 K with high stability (± 0.05 K). An atomic force microscope was used to determine the thickness of the 1$T'$-MoTe$_2$ flake underlying BN. A photograph of a representative 10-nm 1$T'$-MoTe$_2$ sample is shown in Fig. 1(b). In all experiments, films were impulsively excited by 150-fs optical pulses with a central wavelength of 520 nm, focused to about 300 μm full-width at half maximum (fwhm) and with an incident peak fluence of 0.5 mJ cm$^{-2}$. The optical penetration depth of 520-nm light is ~ 30 nm[23]. Note that in addition to acoustic waves, impulsive fs-optical excitation also drives the generation of coherent Raman modes, which are much weaker but still observable in our time-dependent transient absorption traces. For instance, one of such coherent components can be easily discerned as higher frequency oscillations shrouded within the acoustic wave shown in Fig. 1(c). This weaker signal corresponds to the ~ 2.3 THz (~ 77 cm$^{-1}$) Raman modes present in both the $T_d$- and 1$T'$- phases ($^2A_1$ and $^1A_g$ respectively)[24,25]. Transient absorption changes were monitored by time delayed fs-white light pulses generated in a 3-mm YAG crystal and recorded by a dispersive spectrometer. The size of the white light (probe) spot at the sample position was about 20 μm (fwhm). Pump-probe experiments were performed in a quasi-collinear arrangement with a small angle of 10° between the incident pump and probe beams, both linearly p-polarized. The MoTe$_2$ flakes covered the entire area of the freestanding Si$_3$N$_4$ windows. The latter were used to carry out the spatial overlap between the pump and probe beams, which was optimized by maximizing amplitude of the transient absorption signal. The size of the pump beam was much larger than the entire flake thus insuring uniform sample excitation. Owing to our geometrical arrangement and the large size of the pump laser spot, our transient absorption measurements were exclusively sensitive to longitudinal acoustic phonons; i.e. strain waves that propagate in the direction normal to the plane of the layers. An acoustic wave generated by a fs-optical pulse that impinges onto the surface of a thin film is anticipated to propagate normal to the film's surface bouncing back and forth at the film boundaries due to acoustic impedance mismatch; thus yielding a round trip period, $\mathcal{T}_L$, given by Eq. 1,

$$\mathcal{T}_L = 2\,\ell/v_L \qquad (1)$$

where $\ell$ is the thickness of the flake and $v_L$ is the longitudinal sound speed. Note that if such a traveling sound wave evolves into a standing wave, Eq. 1 would still correspond to its fundamental tone. Figure 1(a) shows raw fs-broadband transient absorption spectra obtained for the 10-nm thick $1T'$-MoTe$_2$ sample. There is a clear modulation of the transient absorption signal across the probed spectral range. Figure 1(c) presents the residuals obtained after spectral averaging and removal of the electronic population background dynamics. These steps are performed to improve the signal-to-noise ratio and the confidence of the fitting and Fourier transform procedures. The value of $\mathcal{T}_L = \nu^{-1}$ and its error were calculated by fitting the residuals with a damped sinusoidal function $\mathcal{F}(t) = A\,exp(-\gamma\,t)sin(2\pi\,\nu\,t + \varphi)$. Figure 1(d) also displays the frequency spectrum obtained via fast Fourier transform (FFT) of residuals.

Figure 2 shows the dependence of $\mathcal{T}_L$ with flake thickness. A linear fit of the data with (0,0) intersect renders a value of $v_{L,295K} = 2990 \pm 90$ m s$^{-1}$ at room temperature. This is in reasonable agreement with the reported longitudinal sound velocity for the $2H$-MoTe$_2$ phase, which is $v_L = 3467$ m s$^{-1}$ (exfoliated films)[26], $v_L = 3796$ m s$^{-1}$ (molecular beam epitaxy-grown films)[26] and $v_L = 2800$ m s$^{-1}$ (estimated from DFT calculations)[26,27]. We would like to mention that some discrepancy arising from different layer stacking configurations (i.e. $2H$ versus $1T'$) is expected. However, a relative change of ~ 20% is indeed comparable to the observed variation in the elastic properties of our MoTe$_2$ flakes when the sample temperature is modified by 200 K; as displayed in Fig. 3.

Bulk crystals of 1$T'$-MoTe$_2$ are characterized by a centrosymmetric unit cell with Mo atoms surrounded by Te atoms in a distorted octahedral configuration, slightly shifted from the center and with a $c$-axis inclination angle ($\alpha$) of 93.92º with respect to the plane of the layer. Cooling below a critical temperature of $T_c \sim 250$ K[28] makes 1$T'$-MoTe$_2$ transition into the $T_d$-MoTe$_2$ Weyl semimetal candidate state[29–31], which exhibits a non-centrosymmetric unit cell with $\alpha = 90$º. The inset of Fig. 3 displays the frequency spectrum obtained via FFT for a $\sim 30$-nm thick sample below and above $T_c$. The peak at $\sim 0.38$ THz (13 cm$^{-1}$) corresponds to the characteristic $^1A_1$ interlayer Raman shear mode of the low-temperature $T_d$-phase, which disappears in the high-temperature 1$T'$-MoTe$_2$ state[24,25].

In order to investigate the effect of this first order phase transition on the elastic properties of our exfoliated flakes, we performed fs-transient absorption measurements as a function of the sample temperature. The sound speed is related to the elastic Young's modulus, $E_L$, according to Eq. 2,

$$E_L = \rho\, v_L^2 \qquad (2)$$

where $\rho$ is the density of the material (7.67 g cm$^{-3}$)[32] and $E_L$ has units of pressure. We obtained a value for $E_{L,295\,K} = 68 \pm 4$ GPa. Figure 3 illustrates the dependence of the normalized Young's modulus $E_L/E_{L,295\,K}$ with temperature for two MoTe$_2$ samples with thicknesses of $\sim 12$ nm and $\sim 30$ nm. Given that a typical thermal expansion coefficient of 10$^{-5}$ K would translate into negligible changes of $\ell$ (and $\rho$), it is possible to approximate $E_L/E_{L,295\,K} \approx (v/v_{295\,K})^2$. The measured value of $E_{L,295\,K}$ in 1$T'$-MoTe$_2$ was found to agree reasonably well with the out-of-plane stiffness constant of other layered materials such as graphite (36.5 GPa)[33] and 2$H$-MoTe$_2$ (93.6 GPa, exfoliated flakes)[26]. However, the Young's modulus of MoTe$_2$ displays a much more pronounced temperature dependence, i.e. (–2.0 $\pm$ 0.1) x 10$^{-3}$ K versus –3 x 10$^{-4}$ K for graphite (estimated within the same temperature range from the modeled results from reference [33]). This observation reflects large anharmonicities of the lattice potential along the direction of the weakly interacting van der Waals cohesive forces in this highly anisotropic system[34]. Moreover, the linear behavior illustrated in Fig. 3 suggests that the elastic properties of the films are not substantially altered, neither by the aforementioned $T_d – 1T'$ structural transition nor changes in thickness within the explored $\sim 12$-nm – 30-nm range. The latter statement is also consistent with the linear trend observed in Fig. 1 despite the fact that sub-12-nm thick 1$T'$-MoTe$_2$ films were found to spontaneously transition to the $T_d$-phase by dimensionality effects[22].

We demonstrated a reliable and noninvasive technique that can be used to determine the elastic properties of nm-thick, micron-sized materials, and in particular those of an increasing number of interesting layered transition metal dichalcogenides. Our approach relies on the implementation of homemade nanofabricated optically transparent freestanding Si$_3$N$_4$ windows with dimensions that can be easily adjusted to match the small size of ultrathin flakes typically produced by mechanical exfoliation.

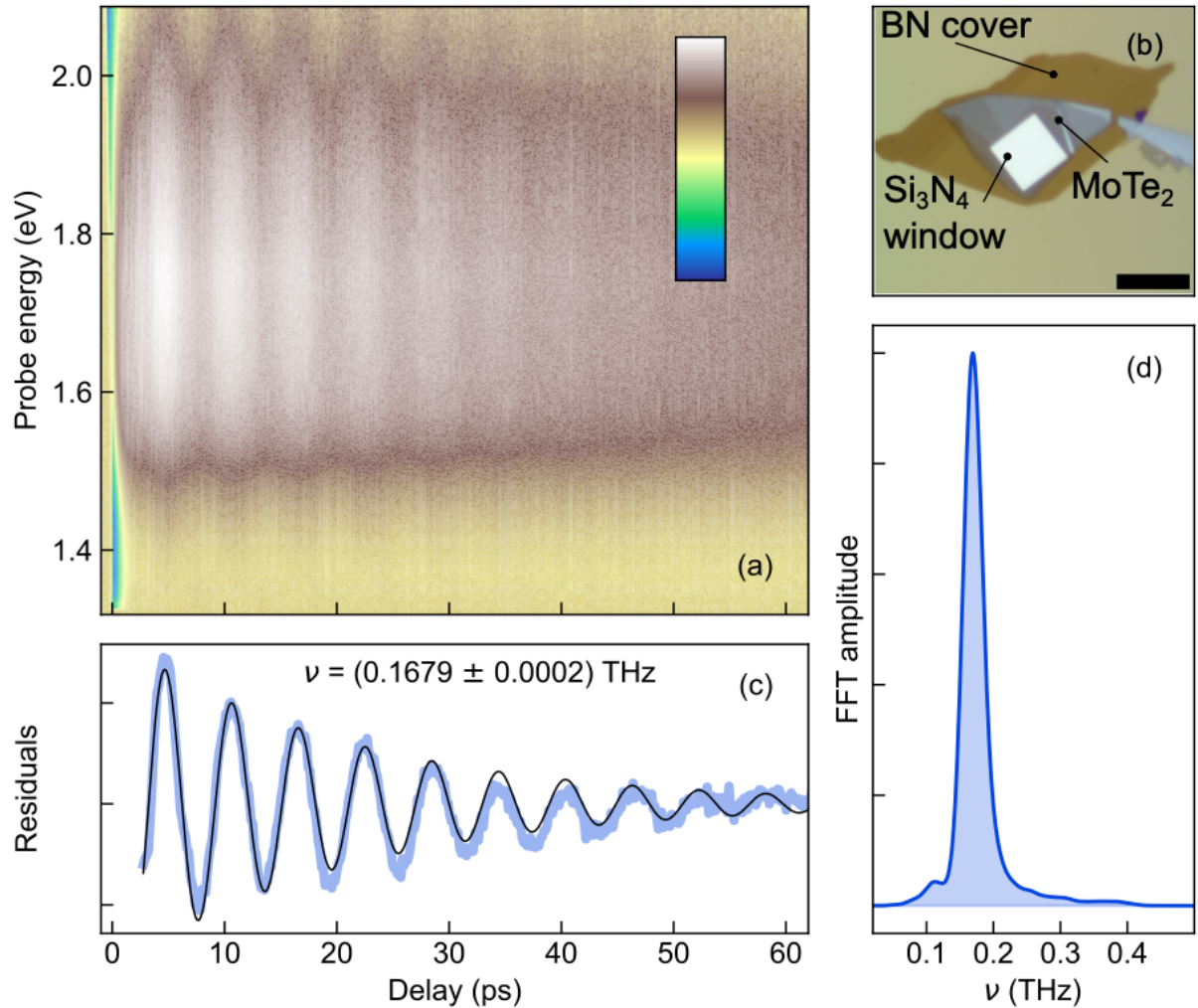

**FIG. 1.** Data analysis scheme illustrated for the study of a 10-nm 1$T'$-MoTe$_2$ film at room temperature, $T$ = 295 K. (a) Raw fs-broadband transient absorption spectra recorded as a function of time delay with a time step of 150 fs. We employed 150-fs pump pulses centered at a wavelength of 520 nm with an incident fluence of 0.5 mJ cm$^{-2}$. (b) Photograph of the prepared flake following transfer and BN capping. The square shape of the freestanding and transparent Si$_3$N$_4$ window is clearly visible. (c) Blue trace: residuals obtained after averaging across the probed photon energy range and removal of the electronic population background dynamics. Black trace: fit of the data by a damped sinusoidal function, see text. (d) Frequency spectrum obtained by FFT of residuals.

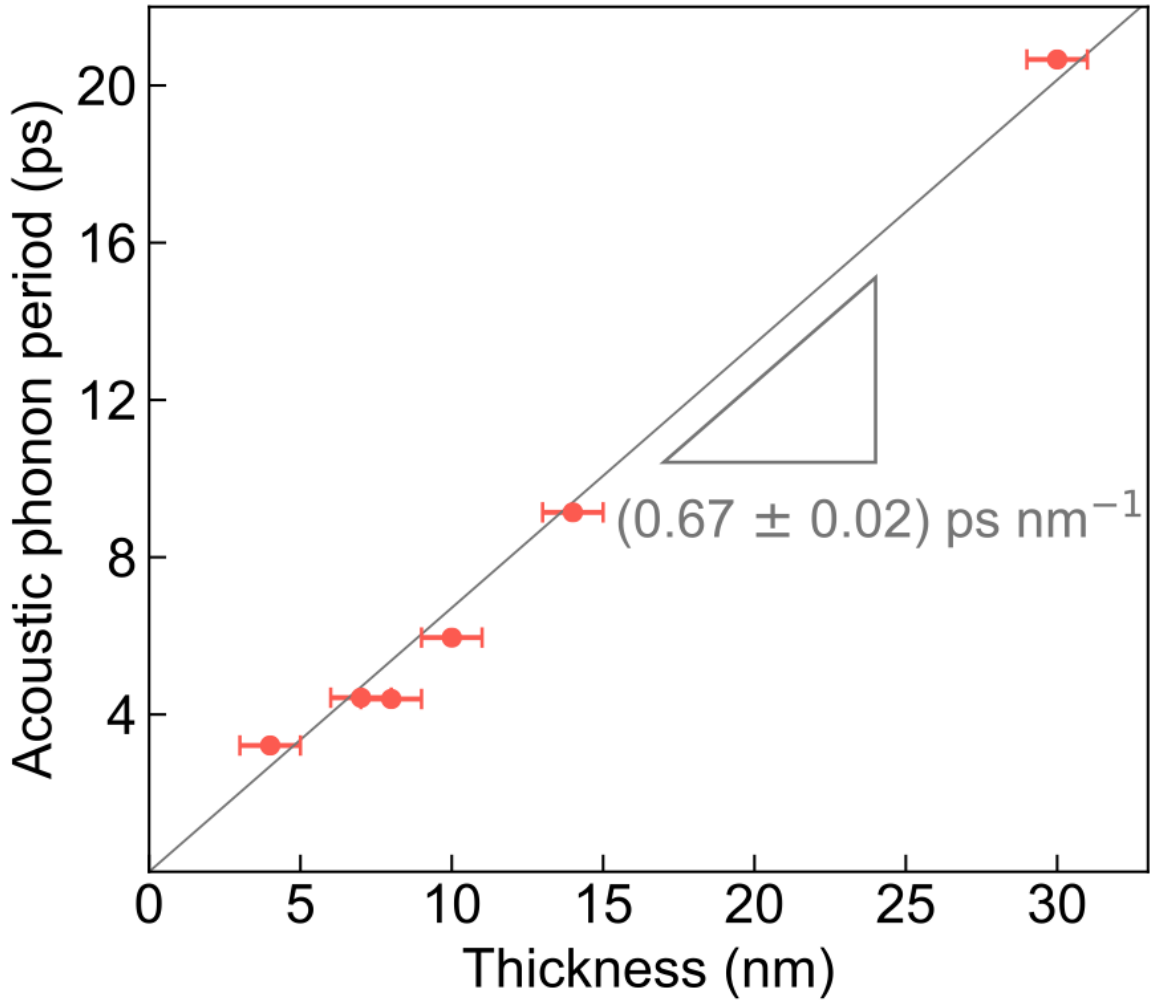

**FIG. 2.** Acoustic wave period extracted from fs-broadband measurements as presented in Fig. 1 for flakes with different thicknesses. The solid trace is a linear fit of the data with (0,0) intersect. According to Eq. 1, the slope (0.67 ± 0.02) ps nm$^{-1}$ provides a value for $v_{L,295K}$ = (2990 ± 90) m s$^{-1}$. The error bars of film thicknesses are ± 1 nm. This is an upper limit that accounts for the fact that the thicknesses of MoTe$_2$ flakes are determined by measuring the changes in height on the BN protective layer which conforms to the features of the underlying MoTe$_2$ film. The error bars of $\mathcal{T}_L$ are smaller than the size of the symbol.

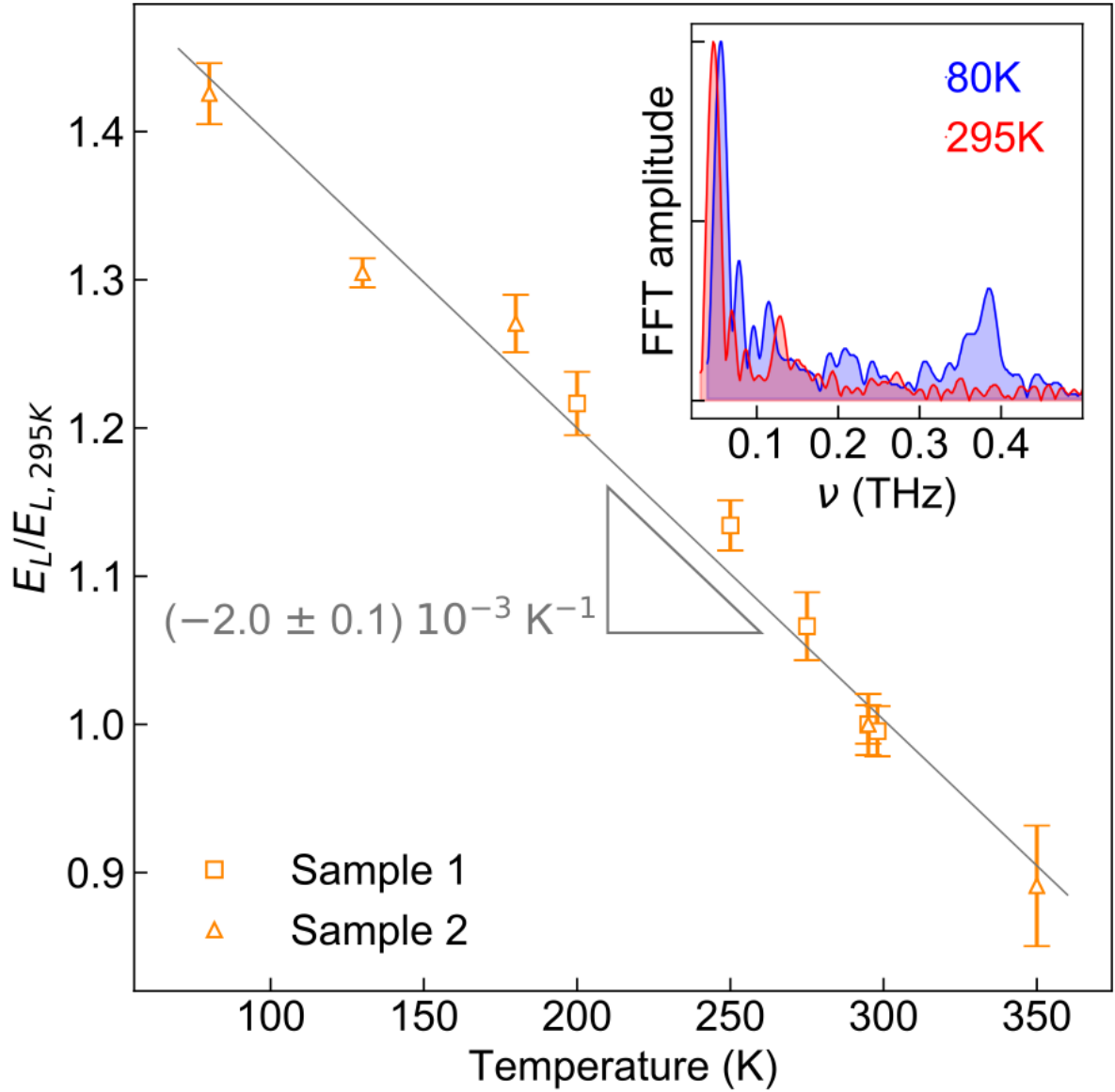

**FIG. 3.** Temperature dependence of the normalized longitudinal Young's modulus approximated as $E_L/E_{L,295\,K} \approx (\nu/\nu_{295\,K})^2$ for two different MoTe$_2$ flakes with thicknesses of ~ 12 nm (sample 1) and ~ 30 nm (sample 2). Note that the $T_d - 1T'$ first order thermal phase transition is expected to occur at $T_c \sim 250$ K. We have not observed any drastic change in elastic properties arising from this phase transition or variations in film thickness in the explored range. Inset: Frequency spectrum obtained via FFT for sample 2 in its $T_d$- and $1T'$-states; blue and red, respectively. The mode at 0.38 THz (13 cm$^{-1}$) corresponds to the characteristic $^1A_1$ Raman interlayer shear mode of the $T_d$-phase.

**ACKNOWLEDGMENTS**

G.S. and A.W.T. acknowledge the support of the National Science and Engineering Research Council of Canada, Canada Foundation for Innovation and Ontario Research Foundation. This research was undertaken thanks in part to funding from the Canada First Research Excellence Fund. G.S. is grateful for the support of the Canada Research Chairs program. G.S. and A.W.T. would like to thank David Cory (IQC) for lending the Oxford cryostat utilized in this work. We acknowledge the support of the Quantum NanoFab (QNF) of the University of Waterloo. F.C., X. L. and Y.S. thank the support of the National Key Research and Development Program under contracts 2016YFA0300404 and the National Natural Science Foundation of China under contracts 11674326,11874357 and the Joint Funds of the National. Natural Science Foundation of China and the Chinese Academy of Sciences' Large-Scale Scientific Facility under contracts U1832141.

**CONFLICTS OF INTEREST**

The authors declare no conflicts of interest.